\documentclass[useAMS,usenatbib]{mn2e}

\title[]{Witnessing the transformation of a quasar host galaxy at z=1.6}
\author[A. Humphrey et al.]{A. Humphrey$^{1,2}$\thanks{E-mail: andrew.humphrey@astro.up.pt}, N. Roche$^{1}$, 
J.~M. Gomes$^{1}$, P. Papaderos$^{1}$, M. Villar-Mart\'{i}n$^{3}$, \newauthor
M.~E. Filho$^{1}$, B.~H.~C. Emonts$^{3}$, I. Aretxaga$^{2}$, L. Binette$^{4}$, B. Oca\~na Flaquer$^{1}$, \newauthor
P. Lagos$^{1}$, J. Torrealba$^{2}$\\
$^{1}$Instituto de Astrof\'{i}sica e Ci\^encias do Espa\c{c}o, Universidade do Porto, CAUP, Rua das Estrelas, PT4150-762 Porto, Portugal\\
$^{2}$Instituto Nacional de Astrof\'{i}sica, \'Optica y Electr\'onica (INAOE), Luis Enrique Erro 1, Sta. Mar\'{i}a Tonantzintla, Puebla, Mexico\\
$^{3}$Centro de Astrobiolog\'{i}a (INTA-CSIC), Ctra de Torrej\'on a Ajalvir, km 4, E-28850 Torrej\'on de Ardoz, Madrid, Spain\\
$^{4}$Instituto de Astronom\'{i}a, Universidad Nacional Aut\'onoma de M\'exico, D.F., Mexico}

\begin{document}

\date{Accepted 2014 November 26. Received 2014 November 25; in original form 2014 May 7}

\pagerange{\pageref{firstpage}--\pageref{lastpage}} \pubyear{2014}

\maketitle

\label{firstpage}

\begin{abstract}
A significant minority of high redshift radio galaxy (HzRG) candidates show extremely red broad band colours and remain undetected in emission lines after optical `discovery' spectroscopy.  In this paper we present deep GTC optical imaging and spectroscopy of one such radio galaxy, 5C 7.245, with the aim of better understanding the nature of these enigmatic objects.  Our g-band image shows no significant emission coincident with the stellar emission of the host galaxy, but does reveal faint emission offset by $\sim$3\arcsec~ (26 kpc) therefrom along a similar position angle to that of the radio jets, reminiscent of the `alignment effect' often seen in the optically luminous HzRGs.  This offset g-band source is also detected in several UV emission lines, giving it a redshift of 1.609, with emission line flux ratios inconsistent with photoionization by young stars or an AGN, but consistent with ionization by fast shocks. Based on its unusual gas geometry, we argue that in 5C 7.245 we are witnessing a rare (or rarely observed) phase in the evolution of quasar hosts when stellar mass assembly, accretion onto the back hole, and powerful feedback activity has eradicated its cold gas from the central $\sim$20 kpc, but is still in the process of cleansing cold gas from its extended halo.

\end{abstract}

\begin{keywords}
galaxies: high-redshift -- galaxies: active -- quasars: general -- quasars: feedback -- galaxies: evolution
\end{keywords}

\section{Introduction}
Despite their rarity, powerful radio galaxies continue to play an important role in cosmological investigations.  They are hosted by massive elliptical galaxies or their progenitors, and can be detected across a vast range in redshift (z$\sim$0 to z$>$5: e.g. van Breugel et al. 1999).  In observing a powerful radio galaxy we get to see a massive galaxy as it undergoes a phase of significant accretion of matter onto its supermassive black hole, the radiative and mechanical output from which can potentially affect the subsequent stellar mass assembly activity and the gaseous properties of the galaxy. As such, radio galaxies are useful probes for understanding the evolution of massive galaxies, and for understanding the possible symbiosis between the build-up of the central black hole and the evolution of its host.  

High-z radio galaxies are often initially identified using their radio properties, with a particularly successful method being to select radio sources with small angular sizes and relatively steep spectra from low frequency radio surveys (e.g. R\"ottgering et al. 1994; Afonso et al. 2011; Ker et al. 2012); optical or near-infrared imaging is then used to identify the likely host galaxy, and detection of rest-frame optical or ultraviolet emission lines using optical or near-infrared spectroscopy is used to determine the redshift of the galaxy. Naturally, this final stage introduces a bias such that radio galaxies which do not have bright emission lines are missing from catalogues of spectroscopically confirmed high-z radio galaxies (e.g. Lacy et al. 1999; De Breuck et al. 2001; Willott, Rawlings \& Blundell 2001 (W01); De Breuck et al. 2006; Bryant et al. 2009; Filho et al. 2011).  Radio galaxies which do not show emission lines, even after long spectroscopic integrations at 4-10 m class telescopes, may account for up to 30 per cent of all steep-spectrum high-z radio galaxy candidates (see e.g. Miley \& De Breuck 2008; Reuland 2005).  In this paper we refer to this class of radio galaxies as 'Line-Dark and Red' radio galaxies (LDRRGs). 

In addition to being undetected in emission lines, LDRRGs also usually have extremely red optical to near-IR colours, implying either heavy reddening or an old stellar population (W01).  They also often have relatively compact radio sources, and typically are bright at millimetre/sub-millimetre wavelengths (Reuland 2005).  LDRRGs may represent a significant phase in the evolution of radio galaxy hosts that has hitherto been mostly overlooked, giving them significant potential to improve our understanding of galaxy evolution.  It has also been suggested that some LDRRGs could be at very high redshifts (i.e., $\ga$5), where the Ly$\alpha$ emission line would be redshifted into spectral regions where sky lines are very strong, thereby rendering the line difficult to detect, or else redshifted out of the optical window altogether; as such, LDRRGs may be considered potential candidates for the earliest radio galaxies in the Universe (e.g. Miley \& De Breuck 2008). Another possibility is that LDRRGs may lie in the so-called `redshift desert' -- a redshift interval corresponding approximately to 1$<$z$<$2, for which none of the brightest UV-optical emission lines fall within the optical wavelength range. In addition, some LDRRGs may be radiatively inefficient radio galaxies, at high-redshift (see e.g. Heckman \& Best 2014, and references therein). 

In this paper we present deep optical imaging and spectroscopy from the 10.4 m Gran Telescopio Canarias (GTC), for the LDRRG 5C 7.245. This is one of seven 7C radio galaxies studied by W01 that are distinguished by extremely red colours (R-K$>$4.1)\footnote{Unless otherwise stated, we use AB magnitudes herein.}, and undetermined redshifts due to an absence of emission lines after $\sim 1$ h of WHT-ISIS optical spectroscopy (Willott et al. 2002). While undetected in the R and I band images of W01 (R$>$24.8), they detected a counterpart in the near infrared (K=20.7). Their United Kingdom Infrared Telescope spectroscopy, also presented by W01, revealed a tentative detection ($\sim$3$\sigma$) of a single emission line at 1.712 $\mu$m, which the authors suggest to be H$\alpha$ at 1.609, and which is broadly consistent with the photometric redshift z$\sim$2.0 they derived by fitting single stellar populations to their broad band RIJHK photometry. On the basis of these properties, we have selected 5C 7.245 to be studied using deep optical imaging and spectroscopy, in a wavelength regime that is expected to yield insights into the emission processes that should dominate near or below the 4000 \AA~ break. With a radio jet power of $\sim$6$\times$10$^{45}$ erg s$^{-1}$ (Punsly 2005; Pen et al. 2009), this radio-loud galaxy is also expected to provide insights into radio-mode feedback in high redshift galaxies. Throughout this work, we assume $H_{0}$=71 km s$^{-1}$ Mpc$^{-1}$, $\Omega_{\Lambda}$=0.73 and $\Omega_{m}$=0.27, which gives a scale of 8.56 kpc arcsec$^{-1}$. 

\begin{figure}
\includegraphics{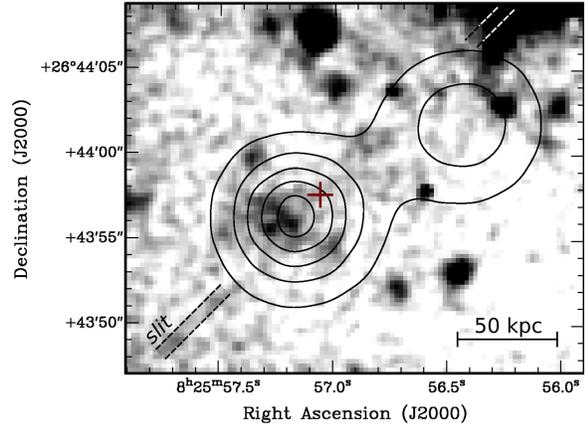}
\vspace{2.27in}
\caption{Smoothed g-band image of 5C 7.245 shown in greyscale, overlaid with contours of the FIRST radio emission. The red cross marks the location of the K-band emission peak from W01. The FIRST contours range from 10 to 90 per cent of the peak flux density (129 mJy beam$^{-1}$). North is up and east is left. The relative astrometry between the radio and the optical coordinate reference frames is expected to be $\sim$0.1\arcsec. }
\label{postage}
\end{figure}

\begin{figure}
\includegraphics{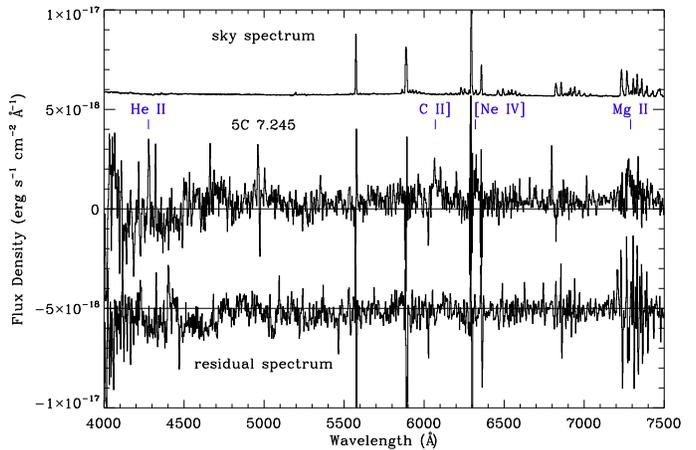}
\vspace{2.3in}
\caption{Spectrum of the SE g-band blob along the radio axis of 5C 7.245, with the four detected emission lines marked.  For comparison we also show a `residual spectrum' extracted from a blank region of sky near to 5C 7.245, and the scaled night sky spectrum, both of which have been shifted by a constant flux density of $\pm$5$\times$10$^{-18}$ erg s$^{-1}$ cm$^{-2}$ \AA$^{-1}$.}
\label{spec}
\end{figure}

\begin{figure*}
\includegraphics{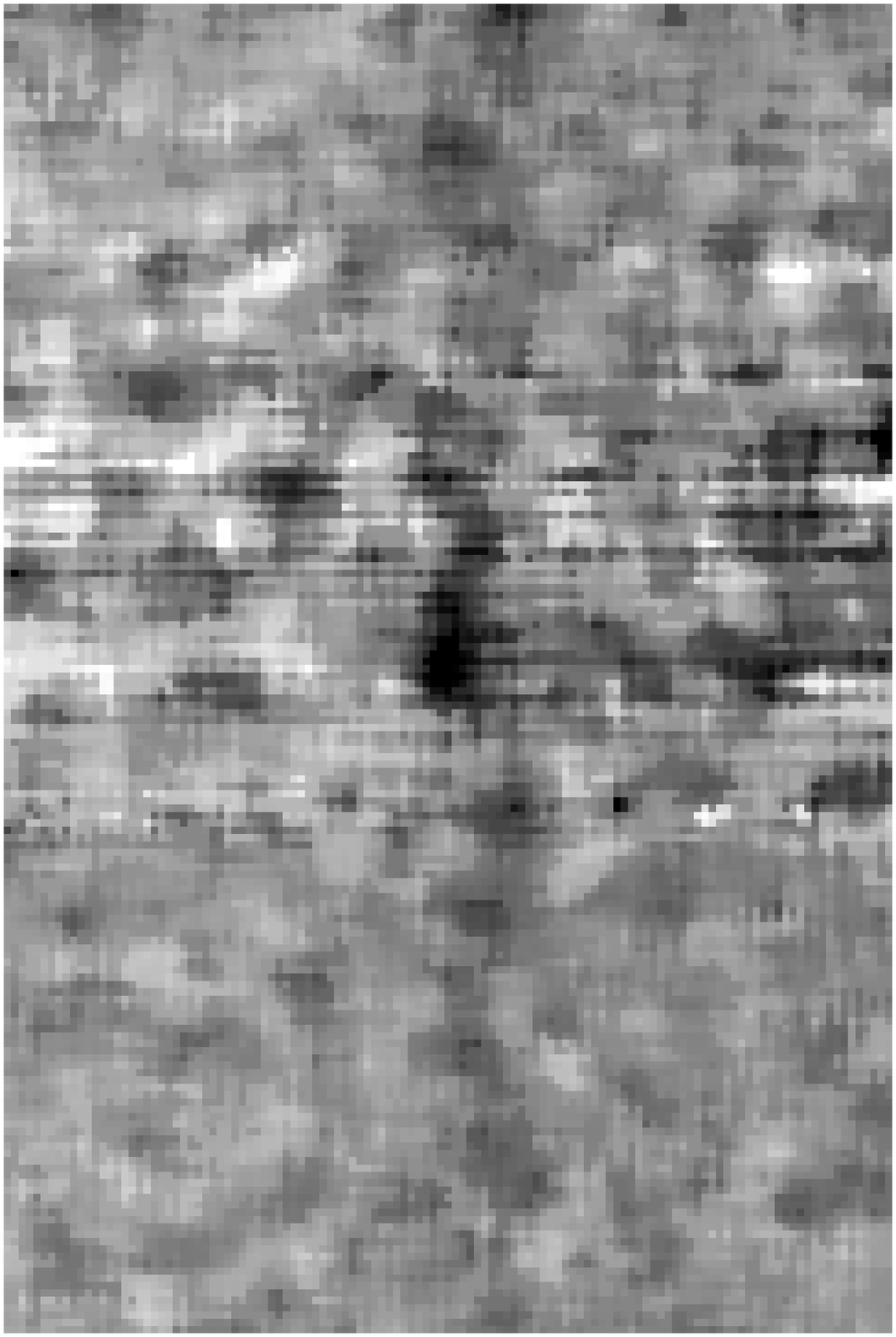}
\includegraphics{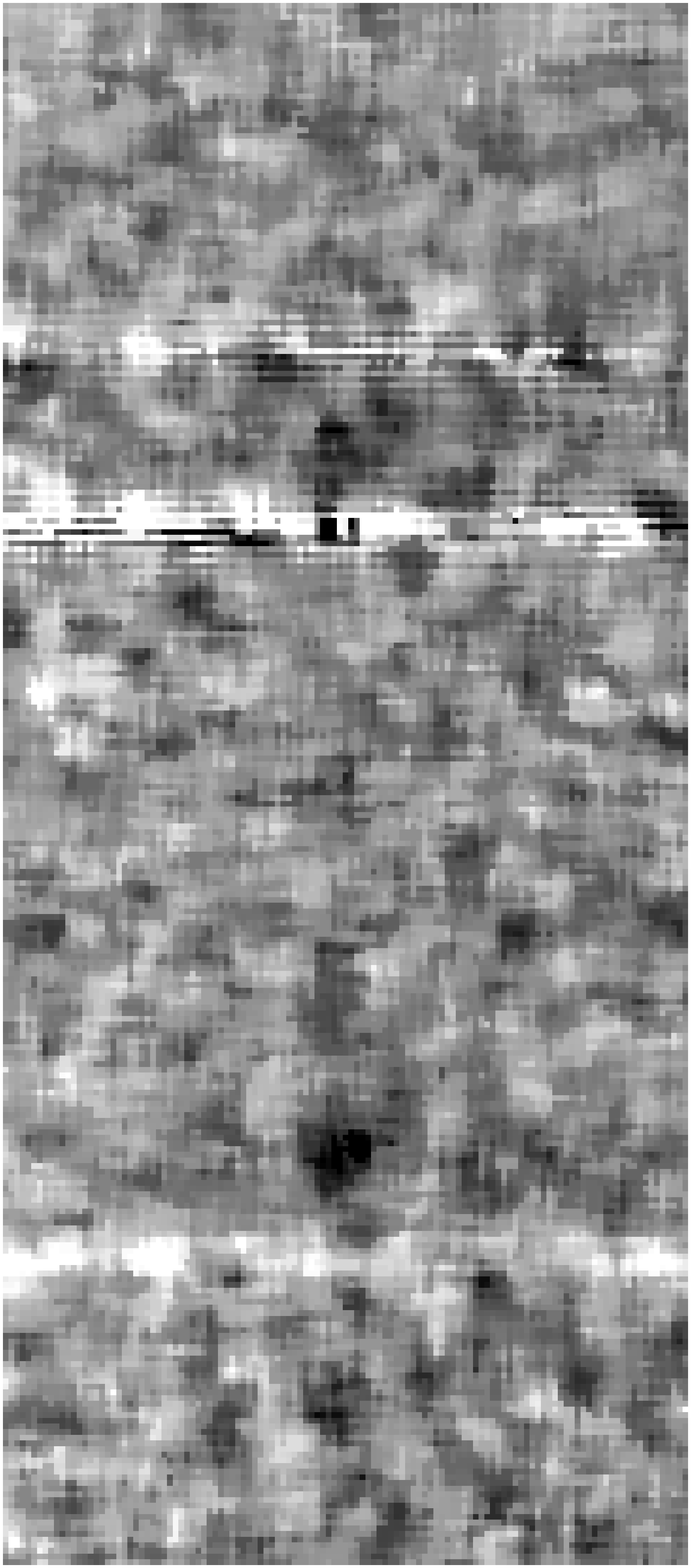}
\includegraphics{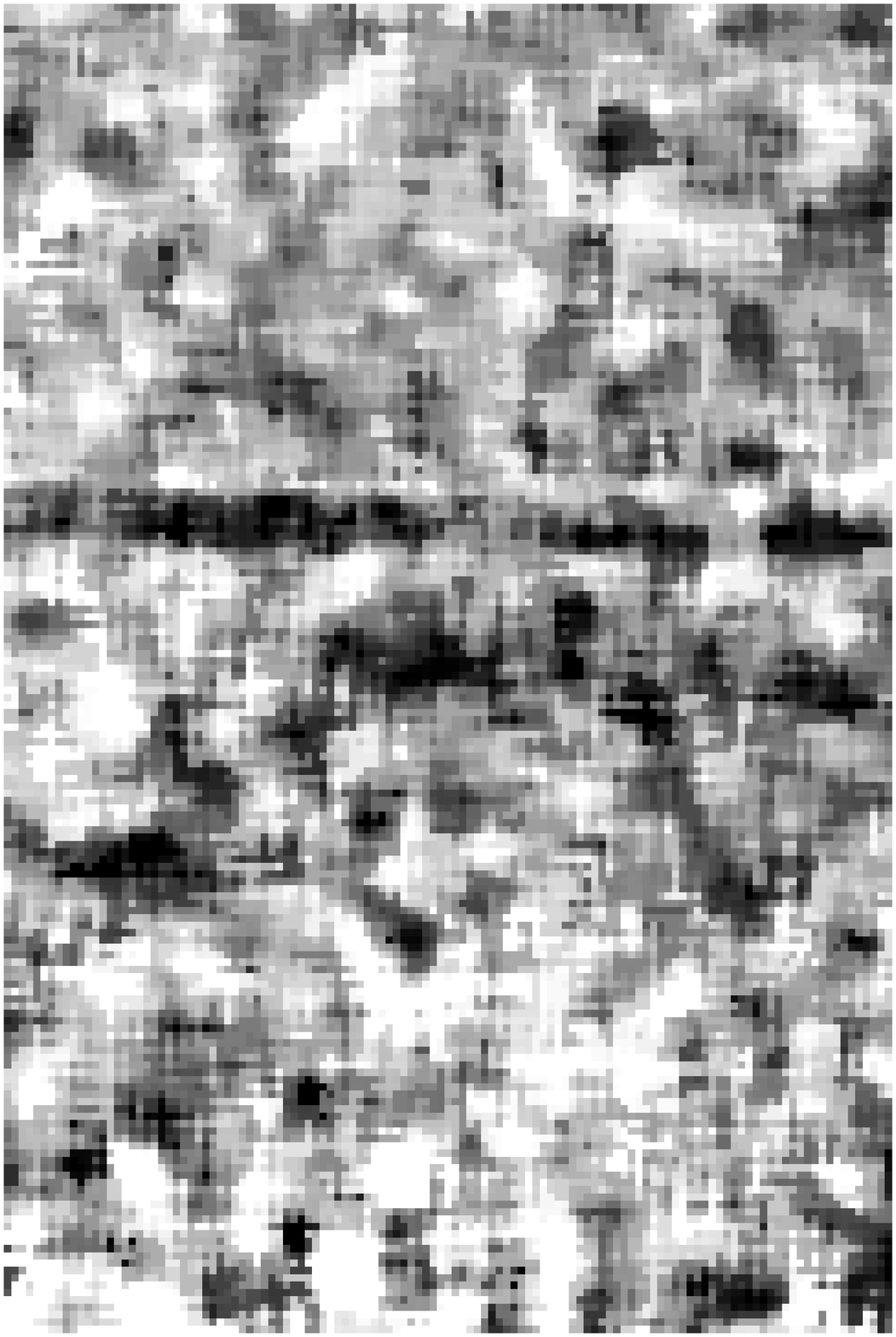}
\vspace{1.2in}
\caption{Sections of the two dimensional spectrum, showing our detections of HeII (left), CII] and [NeIV] (centre) and MgII (right). Wavelength increases leftwards, and the vertical axis is distance along the slit, covering a total spatial extent of 30.7\arcsec~(263 kpc).}
\label{spec}
\end{figure*}

\section{Observations}
The imaging observation was carried out in service mode at the Gran Telescopio Canarias (GTC) on La Palma, Spain, on 28 February 2011.  The image was taken through the Sloan Digital Sky Survey g filter (4050-5580 \AA), using OSIRIS.  Pixel binning of 2$\times$2 was used, resulting in a plate scale of 0.254\arcsec~ per pixel.  The total integration was split into several exposures to assist removal of cosmic rays, bad pixels and ghosts from bright field stars.  The seeing was good, with a full-width at half of maximum (FWHM) of $\sim$1.0\arcsec.  The sky was dark, with clear transparency.  The images were debiased, flat fielded, and then combined, making use of software from the IRAF and STARLINK suites.  Flux calibration was performed using the standard star SA 95-193.  

The long slit spectrum of 5C 7.245 was also obtained in service mode using OSIRIS with GTC, in dark time on 22 January 2012.  The sky transparency was clear, and the seeing FWHM was $\sim$0.9\arcsec.  We used an 0.8\arcsec wide slit with the R1000B grism, which gave a usable spectral range of $\sim$3700-7800 \AA~and an instrumental profile of 6.2 \AA.  The slit was oriented at a position angle of 135$^{\circ}$~(north through east) so as to capture extended emission seen in the g-band image.  The total spectroscopic integration time was 2540 s, split into two individual exposures to aid cosmic ray removal.  The raw data were debiased, flat fielded, wavelength-calibrated and then co-added using standard IRAF procedures, with our flux calibration using observations of G191B2B. Our data were corrected for Galactic extinction using the Schlafly \& Finkbeiner (2011) recalibration of the Schlegel, Finkbeiner \& Davis (1998) dust map.  

\section{Results and Analyses}

\subsection{Optical Image}
\label{phot}
In Fig. ~\ref{postage} we show the greyscaled g-band image of 5C 7.245, with the FIRST (Faint Images of the Radio Sky at Twenty-cm: Becker et al. 1995) radio image overlaid as contours, and with the K-band detection of the radio galaxy host from W01 marked. Although our new optical image shows no significant emission at the position of the peak of the K-band emission (g$\ge$26.2; 3\arcsec~ diameter aperture), we do detect a knot of extended emission located $\sim$3\arcsec~ (26 kpc) South East of the K-band position (g=24.94$\pm$0.09; 3\arcsec~ diameter aperture). Our relative astrometry between the g and K-band images has used field stars present in both images, and has a 1$\sigma$ uncertainty of $\sim$0.3\arcsec. Fig. ~\ref{postage} also reveals the close spatial coincidence between the south eastern radio component and the main g-band structure. The 1$\sigma$ uncertainty between the radio and optical reference frames is roughly $\sim$0.50\arcsec-1\arcsec. 

\subsection{Emission Lines}
During our spectroscopic observations, the slit was placed so as to cover both the K-band position of 5C 7.245 (from W01) and the knot to the SE detected in the g-band image.  Our long slit spectrum shows no emission lines from the K-band position, with a 3$\sigma$ limit of $\sim$3$\times$10$^{-17}$ erg s$^{-1}$ cm$^{-2}$.  However, we have detected HeII $\lambda$1640, CII] $\lambda$2326, [NeIV] $\lambda$2423 and MgII $\lambda$2800 at the position of the g-band knot and the position of the apparent SE radio hotspot (Fig. ~\ref{spec}; Table ~\ref{tab1}).  The flux ratios between the four ultraviolet emission lines appear typical of high-z radio galaxies (see e.g. McCarthy 1993; Humphrey et al. 2008).   

We determine a redshift of z$=$1.6086$\pm$0.0005 from the HeII line, which we expect to provide the cleanest measurement since it is a single, non-resonant line.  The redshifts of the three other emission lines are in agreement with this value, within the uncertainties.  W01 reported a marginal detection of an emission line at 1.712 $\mu$m, suggested to be H$\alpha$ at z$=$1.61, the reality of which we can now confirm.  

\subsection{AGN Location}
\label{location}
Currently available radio images of 5C 7.245 show no clear detection of a flat spectrum radio core, even in the highest spatial resolution radio image (Fig. 1). 5C 7.245 has been detected in X-rays with a flux of 7.1$\pm$2.5$\times$10$^{-14}$ erg s$^{-1}$ cm$^{-2}$ in the 0.2-12 keV band, taken from the XMM-Newton Serendipitous Source Catalogue (2XMM hereafter; Watson et al. 2009). However, the relatively low signal to noise ($\sim$3) of this 2XMM detection results in a positional uncertainty large enough for the X-ray position to coincide with either the g or K-band sources. 

A possible counterpart to 5C 7.245 was detected in the WISE survey. With colours [3.4]-[4.6]=1.4$\pm$0.2 and [4.6]-[12]=3.9$\pm$0.3 (Vega magnitudes), it lies squarely within the `AGN' region of Figure 10 of Yan et al (2013), indicating the presence of an obscured and actively accreting super massive black hole (SMBH). Does this WISE source correspond to either of the g and K-band sources? The WISE detection is at RA 08:25:57.03 and dec +26:45:58.4 (J2000), with a 1$\sigma$ uncertainty of 0.3\arcsec on each of RA and dec, which places it 0.9$\pm$\arcsec NNW of the K-band position of 08:25:57.06 +26:43:57.6 reported by W01. The 2$\sigma$ error in the relative astrometry between the K-band and WISE local coordinate frames is expected to be $\sim$1\arcsec, which allows the K-band and WISE positions to be in agreement within the uncertainties. Given the proximity of the WISE source to the K-band source, and the consistency with an obscured AGN of the colours of the former, we thus conclude that the AGN is located within the K-band source. We also point out that the K-band source has rest-frame optical magnitudes and colours consistent with a passive, giant elliptical galaxy, as is typical for the hosts of powerful, radio-loud active galaxies (see the stellar population fits to the RIJHK magnitudes by W01). 

Could the AGN instead be located within the g-band knot? It cannot, simply because the g-band and WISE sources are far too widely separated ($\sim$4\arcsec) to be considered cospatial. 

\subsection{Excitation: Diagnostic Diagrams}

Interestingly, the position angle defined by the K-band position of the radio galaxy and the SE knot of g-band emission (120$\pm$10$^{\circ}$) is very similar to that of the radio source (120$^{\circ}$).  This is reminiscent of the so-called 'alignment effect', in which optically luminous radio galaxies at intermediate and high redshifts tend to show alignments between the radio-emitting structures and the UV-optical continuum and line emission (Chambers et al. 1987; McCarthy et al. 1987), as a result of illumination by the anisotropic radiation field of the AGN (e.g. Tadhunter et al. 1998), or by the ionization of gas by shocks driven by the radio jets (e.g. Best et al. 2000; Bicknell et al. 2000). Though it is not unusual for powerful radio galaxies to show ionized nebulae extending tens or even a hundred kpc from the AGN (e.g., McCarthy et al. 1995; Reuland et al. 2003), the case of 5C 7.245 is quite unusual in that the spatial peak of the nebular emission does not occur at the position of the AGN (or the peak in the stellar light). It instead resembles detections of cold molecular CO-emitting gas found just beyond the brightest edge of the radio source in several HzRGs (Nesvadba et al 2009, Emonts et al 2014).

It is interesting to consider whether the ionizing radiation field of the central AGN is sufficiently powerful to photoionize the g-band knot at its observed nebular luminosity. We extrapolate the total ionizing photon luminosity of the AGN using the 2XMM X-ray luminosity, and compare against the ionizing luminosity required to explain the observed line emission. Using a spectral energy distribution (SED) and an X-ray to UV spectral slope ($\alpha_{ox}$=-1.37) appropriate to radio-loud AGNs (Miller et al. 2011), and the corresponding X-ray to bolometric correction (Lusso et al. 2012), we estimate a bolometric AGN luminosity of $\sim$1.0 $\times 10^{46}$ erg s$^{-1}$, and an ionizing luminosity of $Q\sim$5.2$\times$10$^{55}$ ph s$^{-1}$. Assuming an ionization bicone opening angle of $\sim$90$^{\circ}$, then the anisotropic ionizing radiation field of the AGN would have a luminosity of $Q_a$$\sim$1.5 $\times$10$^{55}$ ph s$^{-1}$. If all of these ionizing photons were to be absorbed by cold gas in the host galaxy, with 68 \% of the absorber photons resulting in the emission of a Ly$\alpha$ photon, then the total Ly$\alpha$ luminosity would be $\sim$1.6 $\times$10$^{44}$ erg s$^{-1}$, and a HeII $\lambda$1640 luminosity of 1.4 $\times$10$^{43}$ erg s$^{-1}$ if we adopt the typical Ly$\alpha$ / HeII flux ratio of 12 for HzRGs (Humphrey et al. 2008). In 5C 7.245, the observed luminosity in HeII of the g-band knot (8.1$\times$10$^{41}$ erg s$^{-1}$) is a factor of 17 lower than we predict based on the X-ray luminosity, and thus is energetically consistent with photoionization by the central AGN, if there is a relatively high LyC leakage fraction through the host galaxy of 5C 7.245. We note that a high ($>$0.9) LyC escape fraction has been found in the LINER nuclei of several early-type galaxies in the nearby Universe (Papaderos et al. 2013). Even though we recognize that slit losses, due to the transverse size of the g-band knot being larger than the slit width, may have an appreciable impact on this calculation, this is unlikely to account for the discrepancy between expected and observed HeII luminosity. 

\begin{figure}
\includegraphics{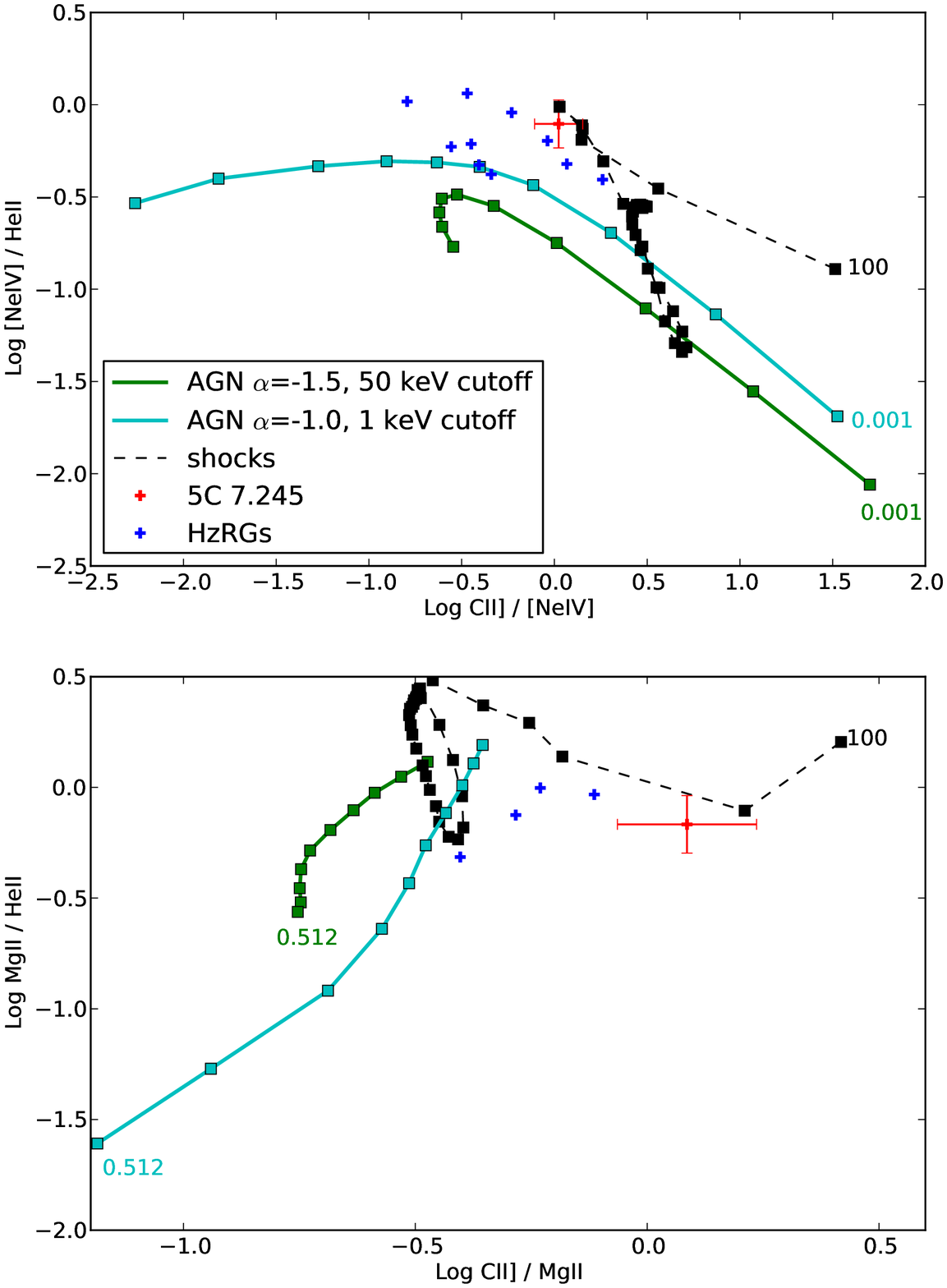}
\includegraphics{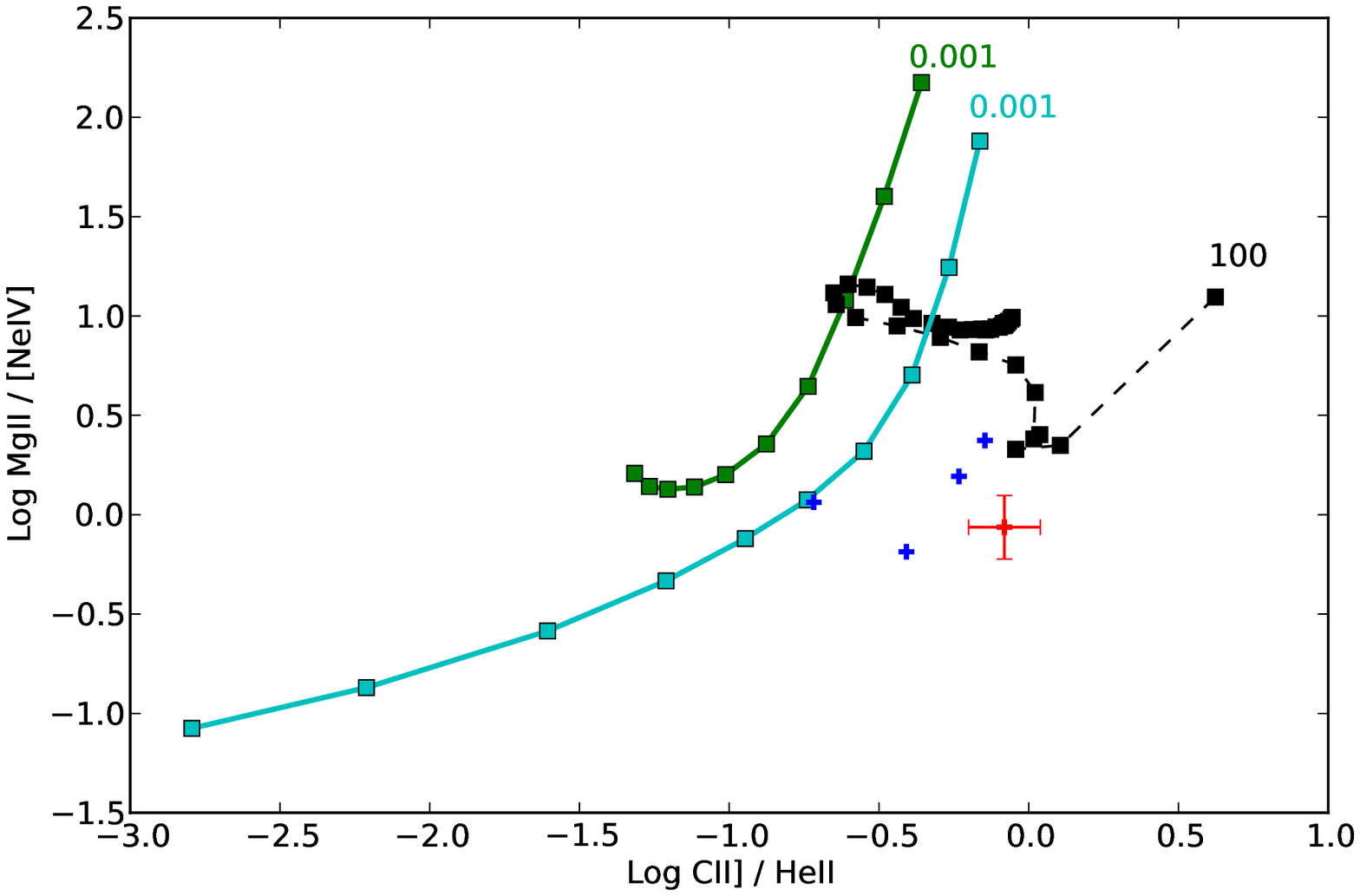}
\vspace{7.5in}
\caption{Diagnostic diagrams involving the UV emission lines HeII, CII], [NeIV] and MgII. We show some of our MAPPINGS 1e photoionization model calculations, together with shock models from Allen et al. (2008), the measured line flux ratios of the g-band knot of 5C 7.245 (this work), and the flux ratios of other HzRGs as reported by Vernet et al. (2001) and Sol\'orzano-I\~narrea et al. (2004). We have labeled one end of each model locus to indicate shock velocity or ionization parameter U as appropriate.}
\label{diagnostic}
\end{figure}

In order to investigate its excitation, in Fig. \ref{diagnostic}~ we show UV emission line ratios measured in the g-band knot, plotted alongside the results of shock-model calculations that combine post-shock cooling gas and its photo-ionized precursor (Allen et al. 2008), and MAPPINGS 1e photoionization code (Binette et al. 2012) calculations appropriate for AGN-photoionized nebulae (e.g. Humphrey et al. 2014). The shock models use a gas density of 100 cm$^{-3}$, solar gas metallicity, magnetic parameter B/n$^{1/2} = e \mu$G cm$^{3/2}$, and a range of shock velocities spanning 100-1000 km s$^{-1}$ in steps of 25 km s$^{-1}$ (see Allen et al. 2008 for further details). Our Mappings 1e photoionization models simulate a cloud of gas, again with density 100 cm$^{-3}$ and solar metallicity, that is being illuminated by a power-law ionizing spectrum of the form $f_v \propto v^{\alpha}$. We have shown two sequences in ionization parameter U, each with a different ionizing spectrum. In order of increasing hardness, these are: (i) $\alpha$=-1.5 with a high energy cut-off at 50 keV (e.g. Robinson et al. 1987; Humphrey et al. 2014) and (ii) $\alpha$=-1.0 with a high energy cut-off at 1 keV (e.g. Villar-Mart\'{i}n et al. 1997; Humphrey et al. 2008). For each of our MAPPINGS 1e model sequences, U occupies the range 0.001-0.512 and increases in steps of factor 2 between each single model calculation. We do not consider photoionization by young stars, because the presence of strong HeII and [NeIV] emission lines, which require a hard spectrum with a substantial flux at energies $>$54 eV, is incompatible with the ionizing SED of a young stellar population. 

As shown in the diagnostic diagrams (Fig. \ref{diagnostic}), the g-band knot lies far from the loci of our MAPPINGS 1e AGN photoionization models, and thus the models fail to adequately reproduce the relative fluxes of the observed UV lines. More specifically, both the CII] and [NeIV] lines are more luminous relative to HeII and MgII than our photoionization models were able to produce. Curiously, 5C 7.245 is not alone in this respect -- the UV line ratios of other HzRGs (Vernet et al. 2001; Sol\'orzano-I\~narrea et al. 2004) appear to show a generally similar discrepancy with the photoionization models. It should be noted that this discrepancy is not due to just one line or line ratio, because it is present in all three diagrams. Of the two photoionization model sequences shown in the diagrams, using a harder ionizing SED does provide a significant, yet insufficient reduction in the discrepancy with the data. 

We have also computed photoionization model sequences using different values of gas density or gas metallicity, or using $\kappa$-distributed electron energies instead of Maxwell-Boltzmann distributed energies, but these additional models provided no significant improvement in our ability to match the observed line ratios, and in some cases exacerbated the discrepancies. To avoid cluttering the diagnostic diagrams, these other models are not shown. 

Conversely, it can be seen from (Fig. \ref{diagnostic}) that the shock model locus touches, or comes close to, the position of the g-band knot of 5C 7.245. In other words, the shock models are better able to reproduce the observed UV line fluxes than our photoionization models, and thus it is reasonable to argue that shocks are the dominant excitation mechanism for this particular subset of UV emission lines, in this particular region of ionized gas. Indeed, the spatial coincidence of the SE radio hotspot with the g-band / emission line knot would be consistent with the presence of fast shocks, presumably being driven into this nebular region by the SE radio jet.

\subsection{Stellar Mass}

The WISE 3.4 $\mu$m band samples the rest-frame emission at $\sim$1.3 $\mu$m, where old stars will dominate the stellar light, and thus can be useful to estimate the stellar mass of the host galaxy. With a WISE magnitude (Vega) of [3.4]=17.4$\pm$0.2, we estimate a stellar mass of $\sim$10$^{11.8}$-10$^{12.5}$ M$_{\odot}$, under the assumption that the rest-frame 1.3 $\mu$m emission is dominated by stars, with our range of mass to light ratio being derived assuming a single burst of age between 10$^8$ and 4.1$\times$10$^{9}$ yr (the age of the Universe at z=1.6). For this we use Bruzual \& Charlot (2003) spectra, a Salpeter (1955) initial mass function, `Padova 1994' evolutionary tracks (Alongi et al. 1993; Bressan et al. 1993; Fagotto et al. 1994a,b,c; Girardi et al. 1996) and six stellar metallicities (0.005, 0.02, 0.2, 0.4, 1.0 and 2.5 times solar). Such a high stellar mass is in excess of any firm estimates for the stellar mass of ultraviolet-luminous radio galaxies at similarly high redshifts made by, e.g., Seymour et al. (2007). Given the WISE colours of this galaxy (see $\S$\ref{location}) it seems likely that its infrared light also has a significant contamination from the obscured AGN itself.

\section{Discussion: Evolutionary Status of 5C 7.245}

In the preceeding analyses we have obtained new insights into the nature of 5C 7.245. We have confirmed that 5C 7.245 is indeed at high-z, but not at the very high redshifts sometimes suggested to be the case for (some) LDDRGs (z$\ge$5; e.g. Miley \& De Breuck 2008).  The large spatial offset between the stellar host (the K-band source) and the warm ionized gas (the SE, g-band knot), and the fact that 5C 7.245 lies in the `redshift desert', are both factors likely to have contributed to past difficulties in detecting emission lines from this radio galaxy (e.g. W01).  

Its optical through infrared colours are consistent with being a massive galaxy at z$=$1.6, whose stellar mass is dominated by old stars (see the stellar population fits by W01), and which harbours a powerful but obscured quasar at its nucleus. Our optical imaging and spectroscopy have revealed the presence of an extended region of warm interstellar gas, which emits continuum, and emission line spectrum consistent with ionization by shocks. This gaseous structure is closely associated with the south-eastern radio hotspot, and is offset by $\sim$3\arcsec~ (26 kpc) from the centre of the host galaxy, where no significant UV continuum or lines have been detected. To find a galaxy with this particular geometry and overall configuration is unusual and, taken at face value, suggests we are witnessing a rare stage in the evolution of quasar host galaxies. 

As our preferred explanation for this ensemble of observed properties, we argue that we are witnessing a quasar in the act of transforming its host galaxy through the elimination of the remainder of its cold gas, by an act of violent, radio-mode feedback, to eventually become a `retired' galaxy (e.g., Stasi{\'n}ska et al. 2008; Cid Fernandes et al. 2010).  This would correspond to a late stage in the life of the AGN, after star formation and accretion onto the SMBH have depleted the host galaxy's reservoir of cold gas, and when feedback activity has eliminated much of the remaining cold gas. In the specific case of 5C 7.245, with little or no ISM in the central region to scatter or reprocess the UV radiation emitted by the AGN, this region is now devoid of UV emission lines and UV continuum, at face value analogous to Lyman continuum leaking galaxies more commonly seen at lower redshift (e.g. Papaderos et al. 2013). However, the elimination of cold gas from 5C 7.245 is visibly incomplete, and a significant reservoir of cold gas still remains within the extended halo of the galaxy, seen in the form of the UV line and continuum emitting g-band knot as a result of heating by processes driven by the AGN. The close spatial association between this knot and the south eastern radio hotspot, taken together with the presence of spectral signatures of fast shocks in the knot, suggests that significant radio-mode feedback activity is still ongoing.

In summary, we suggest that 5C 7.245 is seen in a rare, relatively short-lived evolutionary stage, when feedback has already eliminated much of the cold gas from the central few kpc of the host galaxy, but with the process remaining incomplete and ongoing. 

\section{Other Scenarios}
A potential alternative scenario, albeit less natural, involves heavy extinction by a large-scale screen of dusty material, to explain the red colours and lack of line emission in the central few tens of kpc.  Such circumstances have been suggested to occur during the early stages of luminous AGN activity, during a transition from a relatively obscured star forming galaxy to a powerful, UV-optical active galaxy (e.g. Sanders et al. 1988). However, in the specific case of 5C 7.245, the fact that we are able to detect UV continuum and emission lines appears to be at odds with the idea that this is a very dusty galaxy.

A further alternative invokes a wide-binary AGN system, with a heavily obscured or radiatively inefficient AGN residing within the K-band galaxy, with another obscured but radiatively efficient AGN residing in (and ionizing) the g-band knot. However, not only would this scenario require a fortuitously close alignment between the radio jets and the AGN pair, but it would also require the presence of a UV luminous AGN where no stellar system has been detected at rest-frame optical or infrared wavelengths. 

While neither of these two alternative scenarios can be completely rejected based on the data we have available to us, we do not consider either to provide a particularly natural explanation for the observed properties of the 5C 7.245 system. 

\section{Summary}
Using deep optical imaging and spectroscopy from the GTC, we have investigated the nature and evolutionary status of 5C 7.245, a LDRRG from the sample of W01. We have confirmed spectroscopically its redshift of 1.609. In addition, we have detected a knot of UV continuum and line emission, offset by $\sim$3\arcsec~ (26 kpc) from the peak of the stellar, K-band emission. The UV emission line ratios from this region are consistent with the effects of shock-ionization. However, no UV line or continuum emission has been detected at the peak of the stellar light. 

The geometrical configuration of this galaxy is rather unusual. We suggest that in 5C 7.245 we are witnessing a relatively rare phase in the evolution of quasar host galaxies, where feedback activity has cleansed the central areas of the host galaxy of cold gas, but is still in progress and has not yet completed the elimination of this gas phase from the outer halo regions. 

This work was originally motivated in part by the desire to find the first powerful radio galaxies, at very high redshifts (z$>$5; van Breugel et al. 1999).  While LDRRGs have been proposed as candidates for very high redshift radio galaxies (e.g. Miley \& De Breuck 2008), the results presented herein caution that at least some LDRRGs may in fact be dust-enshrouded or LyC leaking HzRGs at significantly lower redshift.

\begin{table*}
\centering
\caption{Emission line measurements at the position of the South Eastern g-band blob.} 
\begin{tabular}{llllll}
\hline
$\lambda_{obs}$ & Line ID & z & Flux & FWHM & Notes \\
(\AA)                 &            &     & (10$^{-17}$ erg s$^{-1}$ cm$^{-2}$) & (km s$^{-1}$) & \\
\hline
4277.6$\pm$0.3 & HeII $\lambda$1640 & 1.6086$\pm$0.0005 & 4.7$\pm$0.8 & 630$\pm$70 & \\
                            & OIII] $\lambda$1663 &                                  & $\le$5            & 3$\sigma$ upper limit \\
6070                    & CII]                            & 1.610$\pm$0.002 & 3.9$\pm$1.0 & -- & multiplet \\
6320$^{+2}_{-10}$ & [NeIV]                         & 1.608$^{+0.001}_{-0.004}$ & 3.7$\pm$1.1 & -- & partially hidden by sky line residuals \\
7299$\pm$5       & MgII                           &  1.607$\pm$0.001        & 3.2$\pm$1.0 & -- & partially hidden by sky line residuals \\
\hline
\end{tabular}
\label{tab1}
\end{table*}

\section*{Acknowledgments}
AJH, BOF, JMG, PL and PP acknowledge support by the Funda\c{c}\~{a}o para a Ci\^{e}ncia e a Tecnologia (FCT) under project FCOMP-01-0124-FEDER-029170 (Reference FCT PTDC/FIS-AST/3214/2012), funded by FCT-MEC (PIDDAC) and FEDER (COMPETE). AH acknowledges a Marie Curie Fellowship co-funded by the FP7 and the FCT. PP is supported by a 2013 FCT Investigator Grant, funded by FCT/MCTES (Portugal) and POPH/FSE (EC). NR acknowledges the support of Funda\c{c}\~ao para a Ci\^encia e a Tecnologia grant SFRH/BI/52155/2013. BE acknowledges funding through MINECO grant AYA2010-21161-C02-01. MVM's work has been funded with support from the Spanish Ministerio de Econom\'{i}a y Competitividad through the grant AYA2012-32295.LB acknowledges support from CONACyT grant CB-128556. PL is supported by post-doctoral grant SFRH/BPD/72308/2010, funded by the FCT. IA ackowledges support from CONACyT grant CB-2011-01-167291. JT acknowledges CONACyT  basic research grant 151494. JMG acknowledges support from the FCT through the Fellowship SFRH/BPD/66958/2009 and POPH/FSE (EC) by FEDER funding through the program Programa Operacional de Factores de Competitividade - COMPETE. BE acknowledges funding through the European Union FP7 IEF grant Nr. 624351 and MINECO grant AYA2010-21161-C02-01, We also thank Catarina Lobo, and the anonymous referee, for comments and suggestions that helped to improve this manuscript. 

\section*{References}
 
\noindent Afonso J., et al., 2011, ApJ, 743, 122

\noindent Allen M.~G., Groves B.~A., Dopita M.~A., Sutherland R.~S., Kewley L.~J., 2008, ApJS, 178, 20 

\noindent Alongi M., Bertelli G., Bressan A., Chiosi C., Fagotto F., Greggio L., Nasi E., 1993, A\&AS, 97, 851

\noindent Becker R. H., White R. L., Helfand D. J. 1995, ApJ, 450, 559

\noindent Best P.~N., R{\"o}ttgering H.~J.~A., Longair M.~S., 2000, MNRAS, 311, 23

\noindent Bicknell G.~V., Sutherland R.~S., van Breugel W.~J.~M., Dopita M.~A., Dey A., Miley G.~K., 2000, ApJ, 540, 678

\noindent Binette L., Matadamas R., H{\"a}gele G.~F., Nicholls D.~C., Magris C.~G., Pe{\~n}a-Guerrero M.~{\'A}., Morisset C., Rodr{\'{\i}}guez-Gonz{\'a}lez A., 2012, A\&A, 547, AA29 

\noindent Bressan A., Fagotto F., Bertelli G., Chiosi C., 1993, A\&AS, 100, 647

\noindent Bruzual G., Charlot S., 2003, MNRAS, 344, 1000

\noindent Bryant J.~J., Johnston H.~M., Broderick J.~W., Hunstead R.~W., De Breuck C., Gaensler B.~M., 2009, MNRAS, 395, 1099B

\noindent Chambers K.~C., Miley G.~K, van Breugel W., 1987, Nature, 329, 604C
 
\noindent Cid Fernandes R., Stasi{\'n}ska G., Schlickmann M.~S., Mateus A., Vale Asari N., Schoenell W., Sodr{\'e} L., 2010, MNRAS, 403, 1036

\noindent De Breuck C., van Breugel W., R\"ottgering H., Stern D., Miley G., de Vries W., Stanford S.~A., Kurk J., Overzier R., 2001, AJ, 121, 1241D

\noindent De Breuck C., Klamer I., Johnston H., Hunstead R.~W., Bryant J., Rocca-Volmerange B., Sadler E.~M., 2006, MNRAS, 366, 58D

\noindent Emonts B.~H.~C., et al., 2014, MNRAS, 438, 2898

\noindent Fagotto F., Bressan A., Bertelli G., Chiosi C., 1994a, A\&AS, 105, 29

\noindent Fagotto F., Bressan A., Bertelli G., Chiosi C., 1994b, A\&AS, 105, 39

\noindent Fagotto F., Bressan A., Bertelli G., Chiosi C., 1994c, A\&AS, 104, 365

\noindent Filho M. E., Brinchmann J., Lobo C., Ant\'on S., 2011, A\&A, 536A, 35F

\noindent Girardi L., Bressan A., Chiosi C., Bertelli G., Nasi E., 1996, A\&AS, 117, 113

\noindent Heckman T.~M., Best P.~N., 2014, ARA\&A, 52, 589

\noindent Humphrey A., Villar-Mart{\'{\i}}n M., Vernet J., Fosbury R., di Serego Alighieri S., Binette L., 2008, MNRAS, 
383, 11
 
\noindent Humphrey A., Binette L., 2014, MNRAS, 442, 753 

\noindent Ker L.~M., Best P.~N., Rigby E.~E., R\"ottgering H.~J.~A., Gendre M.~A., 2012, MNRAS, 420, 2644K

\noindent Klamer I.~J., Ekers R.~D., Sadler E.~M., Hunstead R.~W., 2004, ApJ, 612, L97

\noindent Lacy M., Rawlings S., Hill G.~J., Bunker A.~J., Ridgway S.~E., Stern D., 1999, MNRAS, 308, 1096L

\noindent Lusso E., et al., 2012, MNRAS 425, 623L

\noindent McCarthy P.~J., van Breugel W., Spinrad H., Djorgovski S., 1987, ApJ, 321L, 29M
 
\noindent McCarthy P.~J., 1993, ARA\&A, 31, 639
 
\noindent McCarthy P.~J., Spinrad H., van Breugel W., 1995, ApJS, 99, 27

\noindent Miley G., De Breuck C., 2008, A\&ARv, 15, 67M

\noindent Miller B.P., Brandt W.N., Schneider D.P., Gibson R.R., Steffen A.T., Wu Jianfeng, 2011, ApJ 726, 20

\noindent Nesvadba N.~P.~H., et al., 2009, MNRAS, 395, L16

\noindent Papaderos P., et al., 2013, A\&A, 555, L1

\noindent Pen U.-L., Chang T.-C., Hirata C.~M., Peterson J.~B., Roy J., Gupta Y., Odegova J., Sigurdson K., 2009, MNRAS, 399, 181 

\noindent Punsly, B., 2005, ApJ, 623L, 9P

\noindent Reuland M., van Breugel W., R\"ottgering H., de Vries W., Stanford S.~A., Dey A., Lacy M., Bland-Hawthorn J., Dopita M., Miley G., 2003, ApJ, 592, 755R

\noindent Reuland M., 2005, PhD Thesis, Leiden Observatory

\noindent Robinson A., Binette L., Fosbury R.~A.~E., Tadhunter C.~N., 1987, MNRAS, 227, 97 

\noindent R\"ottgering H.~J.~A., Lacy M., Miley G.~K., Chambers K.~C., Saunders R., 1994, A\&AS, 108, 79R

\noindent Salpeter E.~E., 1955, ApJ, 121, 161

\noindent Sanders D.~B., Soifer B.~T., Elias J.~H., Madore B.~F., Matthews K., Neugebauer G., Scoville N.~Z., 1988, ApJ, 325, 74S
 
\noindent Schlafly E.~F.,, Finkbeiner D.~P., 2011, ApJ, 737, 103S
 
\noindent Schlegel D.~J., Finkbeiner D.~P. Davis M., 1998, ApJ, 500, 525S

\noindent Seymour N., et al., 2007, ApJS, 171, 353

\noindent Sol{\'o}rzano-I{\~n}arrea C., Best P.~N., R{\"o}ttgering H.~J.~A., Cimatti A., 2004, MNRAS, 351, 997 

\noindent Stasi{\'n}ska G., et al., 2008, MNRAS, 391, L29 

\noindent Tadhunter C.~N., Morganti R., Robinson A., Dickson R., Villar-Martin M., Fosbury R.~A.~E., 1998, MNRAS, 298, 1035

\noindent van Breugel w., De Breuck C., Stanford S.~A., Stern D., R\"ottgering H., Miley G., 1999, ApJ, 518L, 61V
 
\noindent Vernet J., Fosbury R.~A.~E., Villar-Mart{\'{\i}}n M., Cohen M.~H., Cimatti A., di Serego Alighieri S., Goodrich R.~W., 2001, A\&A, 366, 7 

\noindent Villar-Martin M., Tadhunter C., Clark N., 1997, A\&A, 323, 21 

\noindent Watson M.~G., et al., 2009, A\&A, 493, 339

\noindent Willott C.~J., Rawlings S., Blundell K.~M., 2001, MNRAS, 324, 1 (W01)

\noindent Willott C. J., Rawlings S., Blundell K.M., Lacy M., Hill G.J., Scott S.E., 2002, MNRAS 335, 1120

\noindent Yan L., et al., 2013, AJ, 145, 55

\label{lastpage}

\end{document}